\documentclass[a4paper]{jpconf}
\usepackage{graphicx}

\def\aa{\em A\&A}
\def\aaps{\em A\&AS}
\def\ApJ{\em ApJ}
\def\apjl{\em ApJL}
\def\nar{\em NewAR}
\def\pasp{\em PASP}
\def\pasa{\em PASA}
\def\mnras{\em MNRAS}
\def\cqg{\em CQGra}
\def\PRD{\em PhRvD}
\def\RPPh{\em RPPh}

\begin{document}
\title{Electromagnetic follow-up of gravitational wave transient signal candidates}

\author{Marica Branchesi$^{1,2}$ on behalf of LIGO Scientific Collaboration and Virgo Collaboration}

\address{$^1$ DiSBeF, Universit\`a degli Studi di Urbino ``Carlo Bo'', 61029 Urbino, Italy}
\address{$^2$ INFN, Sezione di Firenze, 50019 Sesto Fiorentino, Italy}

\ead{marica.branchesi@uniurb.it}

\begin{abstract}
Pioneering efforts aiming at the development of multi-messenger
gravitational wave (GW) and electromagnetic (EM) astronomy have been
made. An EM observation follow-up program of candidate GW events has
been performed (Dec 17 2009 to Jan 8 2010 and Sep 4 to Oct 20 2010)
during the recent runs of the LIGO and Virgo GW detectors. It involved
ground-based and space EM facilities observing the sky at optical,
X--ray and radio wavelengths. The joint GW/EM observation study
requires the development of specific image analysis procedures able to
discriminate the possible EM counterpart of GW triggers from
contaminant/background events. The paper presents an overview of the
EM follow-up program and the image analysis procedures.
\end{abstract}

\section{Introduction}
The GW detectors, LIGO~\cite{Abb} and Virgo~\cite{Ac}, aim at the
first direct detection of gravitational waves from very energetic
astrophysical events. The most promising sources are mergers of
neutron stars (NS) and/or stellar mass black holes (BH), and the core
collapse of massive stars. These events are also believed to produce
the most electromagnetically luminous objects in the Universe.
Gamma-Ray Bursts (GRBs) are thought to be associated with the
coalescence of NS-NS or NS-BH binaries or the collapse of very massive
stars (see~\cite{Aba1} and references therein).  Another scenario
associated with compact object mergers is the prediction~\cite{Me} of
isotropic optical transients, called \textit{kilonova}, powered by the
radioactive decay of heavy elements synthesized in the merger ejecta.

In this respect, multi-messenger GW and EM astronomy is a very
promising field of research. An electromagnetic counterpart discovered
through a follow-up of a gravitational wave candidate event would
considerably increase the confidence in the astrophysical origin of
the GW signal. The detection of an EM counterpart would give a precise
localization and potentially lead to the identification of the host
galaxy and the redshift. GW and EM observations provide complementary
insights into the progenitor and environment physics.  In the long
term, combined measurements of the source distance and redshift
through GW and EM radiation, respectively, may allow a new way of
estimating cosmological parameters.

The development of a low-latency GW data analysis pipeline has
permitted the use of gravitational wave candidate signals to conduct
the first EM follow-up program \cite{Aba2} (Dec 17 2009 to Jan 8 2010
and Sep 4 to Oct 20 2010) during the last LIGO/Virgo observation
periods.  The present paper summarizes the GW-data analysis followed
to obtain the prompt EM observation, the EM-observation strategy and
the image analysis procedures used to search for the EM-counterparts.
\section{Low latency GW data analysis}
One of the challenges of successfully obtaining prompt EM observations
is to identify the GW candidates quickly: the data from the three
operating detectors (the two LIGOs and Virgo) must be transferred and
analyzed in near-real time.

The LIGO Scientific Collaboration (LSC) and the Virgo Collaboration
developed a low latency GW-data analysis.  Search algorithms
~\cite{Aba3,Se,Bu} run over the data coming from the detectors,
generate a list of triggers and estimate the potential sky position of
the source on the basis of the differences in the signal arrival time
and amplitude at each detector.  Automated procedures are then
designed to select statistically significant triggers suitable for the
EM observation and to determine the telescope pointing positions.
This process, that typically takes $\sim$10 minutes, is followed by a
manual event validation. A team of trained experts is on duty and
evaluate the detector performances.  The entire procedure, from data
acquisition to the alert sent to telescopes, is typically completed
within 30 minutes.

The triggers selected as candidates for EM follow-up are the ones
occurring in simultaneous observations of all three detectors and with
a power above a threshold estimated from the distribution of
background events. A full description of the GW trigger selection and
the entire EM follow-up process is detailed in~\cite{Aba2}.
\section{EM observation strategy and observatories involved in the EM follow-up}
The uncertainty in the source direction reconstruction scales
inversely with the signal-to-noise ratio~\cite{Fa}. GW events near the
detection threshold are localized into regions of tens, or in some
cases even hundreds, of square degrees. Follow-up EM-telescopes with a
wide Field Of View (FOV) are thus required. However, the majority of
EM telescopes have a FOV which is much smaller than the GW angular
error box and additional priors are necessary to improve the location
accuracy and increase the chance that the actual source be EM
observed.

Taking into account the GW detector sensitivity to the signals coming
from NS binaries~\cite{Aba4}, the EM observable Universe is limited to
an horizon of 50 Mpc.  The observation is restricted~\cite{Ka} to the
regions occupied by Globular Clusters and Galaxies (listed in the
Gravitational Wave Galaxy Catalog~\cite{Wh}) within 50 Mpc. To
determine the telescope pointing position, the probability sky map
based on GW data is ``weighted'' taking into account the luminosity
and the distance of these nearby galaxies~\cite{Nu} and globular
clusters. Tens of thousands of galaxies are included within the 50 Mpc
horizon and the GW observable sources will be mainly extragalactic.

The cadence of EM observations is guided by the expected extragalactic
EM counterpart. The optical afterglow of an on-axis GRB peaks few
minutes after the EM/GW prompt emission.  The kilonova model predicts
an optical light curve that peaks a day after the GW event, due to the
time that the out-flowing material takes to become optically thin.
The agreement with the EM facilities allowed several epoch
observations: the first one as soon as possible, the day after and
observations over longer time-lags to cover the transient light curve
dimming.

The agreement with the EM facilities allowed observations as soon as
possible, then the day after the GW event and, repeated observations
over longer time-lag to follow the transient light curve dimming.

The follow-up program involved ground-based and space EM facilities
observing the sky in different EM bands: the {\em Liverpool
Telescope}, the {\em Palomar Transient Factory} (PTF), {\em Pi of
the Sky}, {\em QUEST}, {\em ROTSE III}, {\em SkyMapper}, {\em TAROT}
and the {\em Zadko Telescope} observing the sky in the optical band,
the {\em Swift} satellite with X--ray and UV/Optical telescopes, and
the interferometers {\em LOFAR} and {\em EVLA} in the radio band.
\section{EM image analysis procedures to search for the EM counterpart}
The EM image analysis aims to detect the transient object counterpart
of the GW signal by analyzing a series of images taken in consecutive
epochs after the GW event. The analysis method is conceptually similar
to the one used to study GRB afterglows with one main difference the
arc minute localization of the current generation gamma-ray
observatories allows a significant reduction of the search area with
respect to the GW observations. In the case of a GW event the image
area to analyze is the one occupied by nearby galaxies and globular
clusters in the telescope FOV taking also into account the possible
offset between host center and the transients (observed up to tens of
kpc~\cite{Be} and predicted by simulations up to few Mpc~\cite{Ke}).
Searching for optical transients in a large sky area requires the
development and use of specific image analysis procedures able to
discriminate the EM counterpart from background/contaminant events.

Several analysis pipelines are actually being developed and tested by
the LSC and the Virgo Collaboration in partnership with astronomers.
The image analysis procedure depends on the EM observation band,
however, the main steps in common are: i) the identification of all
transient objects in the images, ii) the removal of contaminating
events (astrophysical background/foreground events and possible fake
transient events linked to the procedure itself).
\subsection{Optical transient search in the wide-field telescope observations}
A total of 14 alerts have been sent out to the telescopes and 9 of
them led to images being taken by at least one optical telescope. The
image analysis procedures able to identify the transients in wide
field optical images are based on two different approaches: the
``image subtraction'' and the ``catalog cross check'' methods.

An example of a pipeline based on ``cross-correlation of object
catalogs'' is the one designed and tested with the images collected by
the two {\em TAROT}~\cite{Kl} and the {\em Zadko}~\cite{Co}
telescopes.  The main steps of the fully automated analysis pipeline
are as follows:
\begin{enumerate}
\item extraction of the catalog of objects visible in the images using
  \texttt{SExtractor}~\cite{Ber};
\item removal of ``known objects'' listed in a star reference catalog
  (USNO-A2.0) by using a positional cross-correlation tool
  \texttt{match}~\cite{Ri}.  A magnitude check is then used to 
  identify from the list of ``known objects'' the ones that show a 
  flux variation with respect to the reference catalog and recover 
  the possible transients overlapping these objects.
\item trace objects in common to several image catalogs by using a
  cross-positional check. This results in a light curve for each traced
  object;
\item rejection of ``rapid contaminating transients'' (like cosmic rays,
  asteroids or noise) by requiring the presence in a number of
  consecutive images;
\item rejection of ``background transients'' by selecting objects lying
  in the image regions associated with the galaxies and globular 
  clusters within 50 Mpc. Each galaxy region takes into account the 
  possible offset between the host galaxy center and the optical 
  transients;
\item rejection of ``contaminating events'' like galaxies, variable stars
  or false transients by analyzing the light curves. The code selects
  the objects that show a luminosity dimming with time. Assuming that
  the dimming is described by a single power-law ${\cal L} \propto
  t^{-\beta}$, corresponding to a linear variation in terms of magnitude
  equal to $m = 2.5 \beta \log_{10}(t) + C$, a ``slope index''
  $2.5\beta$ is defined and evaluated for each objects. The expected
  ``slope index'' for GRB afterglows and kilonova-like light curves is
  around 2.5-3. In practice, a conservative cut is applied by selecting
  as the possible EM counterparts the objects with the ``slope index''
  larger than 0.5. This value has been checked using Monte Carlo
  simulations.
\end{enumerate}
\noindent
For a survey red limiting magnitude of 15.5 and using images assumed
to be observed 1,2 3 days after the GW events, the preliminary results
on the pipeline sensitivity indicate that the majority of GRB
afterglows can be detected further away the GW horizon distance of 50
Mpc, while the kilonova objects can be detected up to a distance of 15
Mpc.  These results are obtained by repeatedly running the pipeline
over sets of {\em TAROT} and {\em Zadko} images where fake on-axis GRB
and kilonova optical transients were injected.
\subsection{X--ray/UV-Optical transient search using the Swift satellite}
The Swift satellite~\cite{Ge} gives the possibility to follow-up in
the X--ray and Ultraviolet/optical bands a transient event detected by
a ground-based observatory. Two candidate GW trigger alerts were sent.
Each alert was observed by Swift twice with a time latency of few
months between the two epoch observations.

The main steps of the X--ray image analysis are: i) the detection of
the source visible in the FOV; ii) the comparison with the number of
serendipitous sources expected in the FOV, statistically estimated
using the 2XMM catalog source counts; iii) the analysis of the light
curve to identify objects that show a flux dimming in the two time
observation images.The following UV/Optical analysis consists of: i)
searching for the counterparts of the X--ray sources detected; ii)
cross-checking with the DSS catalog to identify unknown objects or
objects that show a flux variations, iii) analysis of the light curve
to investigate UV-Optical source variability.
\subsection{Radio Transient Search using LOFAR and EVLA}
The radio observations, in addition to providing information on the
radio emission of the EM counterpart, possibly allow observation of
events not detected at higher frequencies due to obscuration of dust
and daylight that limit the optical sensitivity. Moreover, upcoming
radio arrays and nowadays LOFAR~\cite{St} give the opportunity to
cover multiple sky patches of few tens of sq. degrees (similar to GW
error box) in a single observation.

Five candidate GW-trigger alerts were sent and observed by
LOFAR. LOFAR started recently to explore the low-frequency sky
(30-80/110-240 MHz) never explored before. An automated analysis
procedure to detect transients is still under development. Challenges
to radio transient searches are linked to the noise contaminating
sources from the atmosphere and the time/space varying ionosphere that
complicate the image calibration.

The EVLA~\cite{Pe} started to participate to the low-latency follow-up
project after October 14th when no science GW-trigger were sent to the
telescope. Since the radio afterglow is expected to peak later and
evolve more gradually with respect to the optical and X--ray afterglow
as observed in the GRBs, a high-latency follow-up consisting in three
epochs observations (3 weeks, 5 weeks and 8 months after the GW
trigger) has been performed for two GW trigger alerts. For each of
them the 3 most likely host galaxies were observed. The data analysis
consists of: i) detection of the radio sources in the FOV; ii) light
curve analysis for variability study; iii) identification for
contaminating radio sources, (e.g. the variability of the Radio Active
Galactic Nucleii emission caused by the scintillation of our Galaxy's
interstellar medium).
\section{Concluding remarks}
The present paper reports on the first EM follow-up program to GW
candidates performed by the LIGO/Virgo collaborations together with
partner observatories. Different procedures to detect the EM
counterpart are under test and development for the different EM-bands.
Evaluation of the rate of EM false detections (unrelated to the GW
event and observed by chance in the field) due to astrophysical or
technical contaminants is under study for each analysis procedure and
EM-band.  The follow-up program is a milestone toward the advanced
detector era.  With a ten-fold improvement of sensitivity~\cite{Ab4},
the number of detectable sources increases by a factor of $10^3$.
According to these predictions the advanced detectors will either make
the first detection, or place strict constrains on astrophysically
interesting quantities.  The observation of an EM counterpart could be
a crucial ingredient in deciding the astrophysical nature of the first
event and start GW/EM multi-messenger astronomy.

\end{document}